\documentstyle[epsf,aps,prl]{revtex}
\begin{document}
\draft \twocolumn[\hsize\textwidth\columnwidth\hsize\csname
@twocolumnfalse\endcsname
\title{Effect of quantum noise on Coulomb blockade in normal tunnel
junctions at high voltages}
\author{J.S. Penttil\"a$^1$, \"U. Parts$^1$, P.J. Hakonen$^1$, M.A.
Paalanen$^1$, and E.B. Sonin$^2$}
\address{ $^1$Low Temperature Laboratory, Helsinki University of
Technology,
FIN-02015 HUT, Finland \\
$^2$The Racah Institute of Physics, The Hebrew University of Jerusalem,
Jerusalem 91904, Israel}

\date{\today} \maketitle

\begin{abstract}
We have investigated asymptotic behavior of normal tunnel
junctions at voltages where even the best ohmic environments start to
look like RC transmission lines. In the experiments, this is
manifested by an exceedingly slow approach to the linear behavior
above the Coulomb gap. As expected on the basis of the quantum theory
taking
into account interaction with the environmental modes, better fits are
obtained
using $1/\sqrt{V}$- than $1/V$-dependence for the asymptote. These
results
agree with the horizon picture if the frequency-dependent phase velocity
is
employed instead of the speed of light in order to determine the extent
of
the surroundings seen by the junction.
\end{abstract}
\pacs{PACS numbers: 74.50.+r, 73.23.-b, 73.23.Hk} \bigskip
]
%End of title in one column mode

\section{Introduction}

Coulomb blockade in a single normal tunnel junction is strongly affected
by
the environment if the real part of  impedance is less than the quantum
resistance $R_K=h/e^2 \approx 26$ k$\Omega$. However, a quite general
sum
rule requires that,  at high voltages, the $IV$ curve is of the form
$V=IR +
e/2C$ where $C$ is the geometric capacitance of the tunnel junction
\cite{Ingold}.  The environment strongly influences the way how this
asymptote is approached. Experiments on typical samples with low
resistance
leads exhibit asymptotic $1/V$ tails which can be explained  well using
either the quantum theory of environment \cite{Ingold}, or the
horizon model \cite{GA,WDH}.

In this paper we analyze our experimental results on $IV$ curves for a
variety of single small isolated tunnel junctions, shunted and
unshunted,
with different values of capacitance $C$ and tunneling
resistance $R_T$. Our attention is concentrated especially on the
high-voltage part of
the $IV$ curves obtained for samples with highly
resistive Cr leads. At high frequencies
which are most important for the high-voltage asymptotics such leads
behave
as lossy $RC$ lines. Experimental results are discussed and interpreted
in terms of the quantum theory of environment which predicts
that the high-voltage asymptote is approached as 1/$\sqrt{V}$, {\it
i.e.} much
slower than  $1/V$ tails revealed for a purely real impedance.
Indeed, we find asymptotic $1/\sqrt{V}$ tails experimentally in the
range of voltages 1 -- 10 mV. An important feature of our experimental
conditions is that the junction resistance for some samples is not large
compared with $R_K$
and the strong tunneling corrections have to be taken into account.

We start this paper with a theoretical overview (Sec. II) which
concentrates on predictions of the quantum theory of electromagnetic
environment \cite{Ingold}  for the high-voltage asymptotics of $IV$
curves.
This theory accounts for environmental effects by a phase dependent
factor in
the tunnel Hamiltonian which describes the tunneling rate of electrons
through the junction. Because of the Johnson-Nyquist noise in the
electric
circuit, the phase fluctuates and this affects the tunneling rate. Due
to
exchange of energy between the electron and environmental modes,
the delta-function $\delta(E)$ in energy must
be replaced by a broader distribution function $P(E)$ in the
expression for the tunneling rate. Therefore, this is
called $P(E)$ theory \cite{WDH}. We prefer the term {\em
phase-correlation theory}, introduced in Ref. \cite{CSC},
emphasizing the important role of phase fluctuations and phase memory
in the theory. An alternative {\em voltage-fluctuation} theory suggested
in
Ref. \cite{CSC}, is also discussed. The latter, in contrast to the
phase-correlation theory, predicts an exponential tail for the
high-voltage asymptotics.

The effect of environment is less pronounced in the high-voltage
than in the low-voltage part of the $IV$ curve, and therefore it is more
difficult for scrutiny. But we show in the theoretical overview that the
high-voltage asymptotics is governed by only {\em the quantum part} of
the
Johnson-Nyquist noise, and therefore is quite important for a reliable
comparison with the quantum theory. The high-voltage asymptotic is also
convenient for studying strong tunneling corrections which are
important at our experimental conditions. In Sec. II we argue that a
proper account of  strong tunneling corrections is to include the
junction resistance as a lumped element in an effective electric circuit
used
for calculation of the Johnson-Nyquist noise. This view is proven by
comparing this approach with a more elaborate theoretical analysis
\cite{O,GZ}.

The experiment and its comparison with the theoretical predictions is
described in Sec. III. We fit the experimental high-voltage tails by a
combination of $1/{V}$ and  $1/\sqrt{V}$ tails with their amplitudes as
fitting parameters, and compare the outcome with the values
calculated for our effective electric circuit. Note that experimentally
it
is  rather hard to distinguish between different power-law dependences
on
voltage unless separately determined parameters are employed in
restricting
the fitted formulas. But there is a clear difference between the
power-law and
exponential tails, and our fitting is in favor of the
power-law tails predicted by the phase-correlation theory. The paper is
concluded by discussion (Sec. IV).

\section{The $IV$ curve asymptotics}

\subsection{The phase-correlation theory}

The $IV$ curve is given by
\begin{equation}
  I= { e}[\Gamma^+(V)-\Gamma^-(V)] ~,
        \end{equation}
where the forward tunneling probability reads \cite{Ingold}
\begin{eqnarray}
\Gamma^+(V) = {1\over e^2 R_T}\int_{-\infty}^\infty dE
\int_{-\infty}^\infty
dE' f(E)[1-f(E')]  \nonumber \\
\times \int_{-\infty}^\infty {dt \over 2\pi \hbar}
\exp\left[\frac{it}{\hbar}(E-E' +eV)\right]
\langle e^{i \varphi(t)} e^{-i \varphi(0)} \rangle~,
    \label{phase} \end{eqnarray}
and the backward tunneling rate $\Gamma^- (V)=\Gamma^+(-V)$.
Here $f(E)$ is the Fermi distribution, and $\varphi(t)
={e \over  \hbar} \int_{-\infty}^t dt^* \delta V(t^*)$,
is the fluctuation
of the phase difference due to fluctuating voltage $\delta V$ across the
junction, which is treated as a
quantum-mechanical operator. The averaging $\langle ... \rangle$ is
performed over possible states of quantum environment, {\it i.e.}~an
ensemble
of modes in an electric circuit which the junction is embedded in. If
the
phase does not fluctuate, the integral $\int dt$ in Eq. (\ref{phase})
yields
the energy delta-function $\delta(E-E'+eV)$, and Eq. (\ref{phase})
reduces
to the usual expression for an ohmic  tunnel junction. But taking into
account the
phase fluctuations,   Eq. (\ref{phase}) yields after some algebra (see
Eq.
(56) in Ref. \cite{Ingold}):
\begin{equation}
\Gamma^+(V) = {1\over e^2 R_T}\int_{-\infty}^\infty dE \frac{E}{1
-\exp\left(-\frac{E}{k_BT}\right)}P(eV-E)~.
    \label{ProbE} \end{equation}
Here
\begin{equation}
P(E)={1\over 2\pi \hbar} \int_{-\infty}^\infty dt \exp\left[ J(t) +
\frac{iEt}{\hbar}\right]~,
        \end{equation}
and
\begin{eqnarray}
J(t) &=& \langle[ \varphi(t) -\varphi(0)]\varphi(0)\rangle
\nonumber \\
&=& 2\int_{-\infty}^\infty {d\omega \over \omega}
\frac{\mbox{Re}  Z(\omega)}{R_K}
 \frac{e^{-i\omega t} -1}{1-e^{-\beta \hbar \omega}}
      \end{eqnarray}
is the phase-correlation function where $\beta=1/k_B T$.
Since $\mbox{Re} Z(\omega)$ is an even function of
$\omega$, the imaginary  part
of $J(t)$ {\em does not depend on temperature}:
\begin{equation}
\mbox{Im}J(t)= -\int_{-\infty}^\infty {d\omega \over \omega}
\frac{\mbox{Re}
Z(\omega)}{R_K}\sin (\omega t)~.
        \label{ImJ} \end{equation}
Using the time-domain formulation \cite{OFS,JED} one can rewrite Eq.
(\ref{ProbE}):
\begin{equation}
\Gamma^+(V) = {1\over2\pi \hbar e^2 R_T}\int_{-\infty}^\infty dt
\gamma(t)\exp\left[ J(t) + \frac{ieVt}{\hbar}\right]   ~,
    \label{ProbT}\end{equation}
where
\begin{eqnarray}
\gamma(t) &=& \int_{-\infty}^\infty dE
\frac{E}{1-\exp\left(-\frac{E}{k_BT}\right)}
\exp\left( - \frac{iEt}{\hbar}\right) \nonumber \\
&=& i\pi\hbar ^2 {d\over dt} \delta(t)
- {\pi^2 \over \beta^2} \frac{1}{\sinh ^2 (\pi t/ \hbar \beta)}~.
   \end{eqnarray}
Then using the sum rules $J(0)=0$ and $i J'(0)=e^2/2C\hbar$
\cite{Ingold}, the current becomes
\begin{eqnarray}
  I = {1\over R_T} \left[ V -{e \over 2C} +
{\pi \over e\hbar \beta^2} \right. \nonumber \\ \left.
 \times   \int_{-\infty}^\infty \frac{dt}{\sinh ^2 (\pi t/ \hbar \beta)}
\mbox{Im}
\{e^{J(t)}\}\sin \frac{eVt}{\hbar} \right] ~.
        \end{eqnarray}
The effect of the environment is given by the last term. The
second derivative of the current at high
$V \gg \pi /e\beta=\pi k_BT /e$ becomes
\begin{equation}
  {d^2I \over dV^2}= -{e\over \pi \hbar R_T}  \int_{-\infty}^\infty
dt \mbox{Im}
\{e^{J(t)}\}\sin \frac{eVt}{\hbar}.
          \end{equation}

The asymptotic behavior at $V \rightarrow \infty$ is determined by the
short-time expansion of the correlation function $J(t)$ where it is
small.
Using Eq. (\ref{ImJ}) we obtain
\begin{eqnarray}
  {d^2I \over dV^2}&=& -{e\over \pi \hbar R_T}  \int_{-\infty}^\infty
dt  \mbox{Im} \{J(t)\}\sin \frac{eVt}{\hbar} \nonumber \\
&=& {e\over \pi \hbar R_T}  \int_{-\infty}^\infty
dt  \int_{-\infty}^\infty {d\omega \over \omega} \frac{\mbox{Re}
Z(\omega)}{R_K} \sin (\omega t)\sin \frac{eVt}{\hbar} \nonumber \\
&=&{2 \over R_T R_K} \frac{\mbox{Re}
Z(eV/\hbar)}{V}.
    \label{Z-ohm}      \end{eqnarray}
The total impedance of the circuit may be presented as
$Z^{-1}(\omega)=Y_0
(\omega)- i\omega C$, where the admittance $Y_0(\omega)$ refers to the
whole circuit except for the capacitive channel of the tunnel
junction.  At high frequencies (voltages) one has:
\begin{equation}
 {d^2I \over dV^2} \approx  {e^2 R_K\over 2\pi^2 C^2 R_T}
\frac{\mbox{Re}Y_0(eV/\hbar)}{V^3}~.
   \label{Z-om1}  \end{equation}

Now integrating twice from $V$ to $\infty$ we obtain the $IV$ curve
\begin{equation}
I={1 \over R_T} \left[V- {e\over 2C} + V_t(V)\right]~,
    \label{IV-g}  \end{equation}
where the ``tail'' voltage is
\begin{equation}
V_t={e^2 R_K \over 2\pi^2 C^2} \int_V^\infty dV_1 \int_{V_1}^\infty dV_2
\frac{\mbox{Re}Y_0(eV_2/\hbar)}{V_2^3}~.
  \end{equation}
Alternatively, the effect of the tail can be presented as a voltage
dependent
correction to the junction capacitance:
\begin{equation}
I = {1
\over R_T}\left(V - {e \over 2 \tilde C(V)}\right)
 \label{IVc}  \end{equation}
where the voltage dependent capacitance $\tilde C$ is
\begin{equation}
\tilde C(V) =C\left(1 + {C \over e}V_t(V)\right)  ~.
   \label{ohmic} \end{equation}
In the ohmic case, $Y_0=R^{-1}$ and
\begin{equation}
V_t =  \frac{R_K}{R} \left(\frac{e}{2\pi C }\right)^2 {1\over V} ~.
 \label{IV-ohm}  \end{equation}
Then Eq. (\ref{IV-g})  agrees with Eq. (115) of Ref. \cite{Ingold}.

Thus the $IV$ curve at high voltages scans the high-frequency impedance
and
does not depend on temperature. This means that only the quantum noise
of
environment affects the high-voltage behavior, and this is manifested by
proportionality of high-voltage tails to the quantum resistance $R_K$.
But
one must remember that the derived asymptotic behavior starts at
voltages $V$
higher than $\pi k_B T/e$. This could make the observation of asymptotic
tails
impossible at very high temperatures because of their very small
contribution which could be masked by nonlinear corrections to the
junction
conductance at very high voltages.

\subsection{Alternative theory of the environment effect:
voltage-fluctuation
theory}

There is another approach to take into account environmental
modes on
the tunneling rate. One uses the expression for the tunneling
probability
obtained for a {\em static} voltage at the junction assuming that this
voltage is a random quantity described by the Gaussian
distribution
arising from the Johnson-Nyquist noise. Let us call it the
voltage-fluctuation theory. Initiated by Cleland {\sl et
al.} \cite{CSC} such an approach was considered as a heuristic model, in
contrast to ``the more accurate'' phase-correlation model. We
argue that the two approaches  are not at all equivalent when compared
with
each other. They are {\em essentially different} in the physical picture
of
the phenomenon. The phase-correlation theory assumes that phase
fluctuations
affect the tunneling probability via a phase-dependent factor in the
tunneling Hamiltonian. This phenomenon represents an example of a
``phase-memory effect''.  In contrast, the voltage-fluctuation theory
assumes
that the tunneling probability depends only on the  voltage at the
present
moment, without any memory effect. But the voltage fluctuates due to the
Johnson-Nyquist noise in the circuit and this affects the current
through the
junction.

In spite of completely different starting assumptions, the results of
the
voltage-fluctuation and the phase-correlation theories agree on certain
aspects. In particular, both the theories demonstrate that quantum
fluctuations of the environment become important when the
circuit resistance $R$ becomes of the order of the quantum resistance
$R_K$.
However, there is an essential difference in predictions for the
following effect:
At low voltages when the Coulomb effects are the most pronounced, the
voltage-fluctuation theory predicts an ohmic behavior for zero-bias
anomaly, but
with an exponentially large resistance value compared with the nominal
tunneling
resistance $R_T$.
In contrast, the phase-correlation theory \cite{Ingold} gives a
nonanalytic
$IV$ curve with a non-linear power law. At high voltages the
voltage-fluctuation theory predicts an exponentially decreasing
quantum-fluctuation correction  to the $IV$ curve, against a power-law
decrease in the phase-correlation theory as derived above.

These two theories were compared in experiments by Farhangfar
{\em et
al.} \cite{Pec} at low voltages, and this comparison was in favor of the
phase-correlation
theory. The present paper addresses the $IV$ curve at high-voltages and
we
shall discuss predictions of the voltage-fluctuation theory for the
asymptotics of the  $IV$ curve. Forward and backward tunneling
probabilities
at zero temperature are
\begin{equation}
\Gamma^\pm(V) = {1\over eR} \int_{{e\over 2C} \mp V}^\infty  \left(\pm
V+\delta V
-{e\over 2C}\right)  p(\delta V)\, d\delta V~.
     \label{W} \end{equation}
Here
$p(\delta V)$ is the voltage fluctuation probability distribution which
we
assume to be Gaussian:
\begin{equation}
p(\delta V)=\frac{1}{\sqrt{2\pi} \Delta V}\exp \left(-\frac{\delta
V^2}{2
\Delta V^2} \right)~,
  \label{P} \end{equation}
and $\Delta V=\sqrt{\langle\delta V^2\rangle}$ is the standard deviation
of this
distribution. Then after integration in the limit $V \gg e/2C,
\Delta V$ ($\Gamma^-$  is insignificant in this limit):
\begin{equation}
\Gamma^+={1\over eR}\left\{ V-{e\over 2C} +
\frac{\Delta V^3}{\sqrt{2\pi}V^2} \exp
\left[-\frac{V^2}{2 \Delta V^2} \right]\right\} ~.
   \label{W+} \end{equation}
This yields the exponential asymptotic voltage tail
\begin{equation}
   V_t = \frac{\Delta V^3}{\sqrt{2\pi} V^2} \exp
\left(-\frac{ V^2}{2 \Delta V^2} \right)~.
    \label{IV-l} \end{equation}

Similar to phase fluctuations, voltage fluctuations are determined by
the Johnson-Nyquist noise in the circuit and
\begin{equation}
\Delta V^2  ={\hbar ^2 \over e^2} \langle \dot \varphi(0)^2\rangle
= {\hbar \over \pi}\int_{-\infty}^\infty \omega d\omega
 \frac{\mbox{Re}  Z(\omega)}{1-e^{-\beta \hbar \omega}}~.
    \label{disV}   \end{equation}
Thus, in contrast to the phase-correlation theory, the high-voltage
asymptote is affected not by a single noise mode with frequency $\omega
=eV/\hbar $, but by the whole noise spectrum. As a result, the voltage
tail must decrease exponentially, which is not confirmed in our
experiments (see Sec. \ref{Exp}). In the rest of this section we
restrict ourselves with discussion of the phase-correlation theory.

\subsection{Strong tunneling effects}

Originally, the phase-correlation theory was developed for the weak
tunneling regime, where the junction resistance $R_T$ is large (
$\gg R_K$) and its
contribution to the equilibrium noise in the circuit is negligible. In
our experiment $R_T \sim R_K$ and we must include also the
strong tunneling effects. Usually tunneling is defined as strong
when the junction resistance becomes of the order or smaller than $R_K$.
But in fact one must take into account strong tunneling
effects even if $R_T$ becomes comparable with the circuit resistance.
Let
us consider the case when $R_T$ is the smallest of all parallel
resistors
in the circuit. Then $Y_0 \approx 1/R_T$ and
\begin{equation}
I = {1\over R_T} \left[
V -{e \over 2C}\left(1 -  \frac{R_K}{R_T} \frac{e}{2\pi^2 C } {1\over
V}\right) \right] ~.
 \label{IV-T}  \end{equation}
This asymptotic expression for the $IV$ curve looks like an expansion in
the
parameter $\alpha e/CV$ where $\alpha =R_K/R_T$ characterizes
dissipation.
The same parameter determined strong tunneling effects analyzed by
Odintsov \cite{O} (see also the expression for the $IV$ curve and
discussion in Sec. 5.3.4 of Ref.
\cite{SZ}). Expanding the expression  for the $IV$ curve of Odintsov in
$\alpha e/CV$,  the first terms of the expansion exactly reproduce Eq.
(\ref{IV-T}) derived as an asymptotic expression for high voltages. The
results by Odintsov  \cite{O} have also been confirmed by a more general
analysis
of strong tunneling by Golubev and Zaikin \cite{GZ} using path-integral
technique.

Thus, at least at high voltages, strong tunneling effects can be
taken into account simply by including the tunnel resistance $R_T$ into
the effective electric circuit. This means that $R_T$ contributes to
the equilibrium Johnson-Nyquist noise on equal terms with other
dissipative
elements of the circuit. The same approach was used
by Joyez {\sl et al.} \cite{JED} in the analysis of the
low-voltage part of the $IV$ curve, but instead of using $R_T$
as an element of the
electric circuit to account for the noise from the junction, they used the
differential resistance at zero-voltage bias, which may be much larger
than $R_T$.

\subsection{Circuit elements as transmission lines}

\begin{figure}
\begin{center}
\epsfxsize=70mm \epsffile[133 137 410 536]{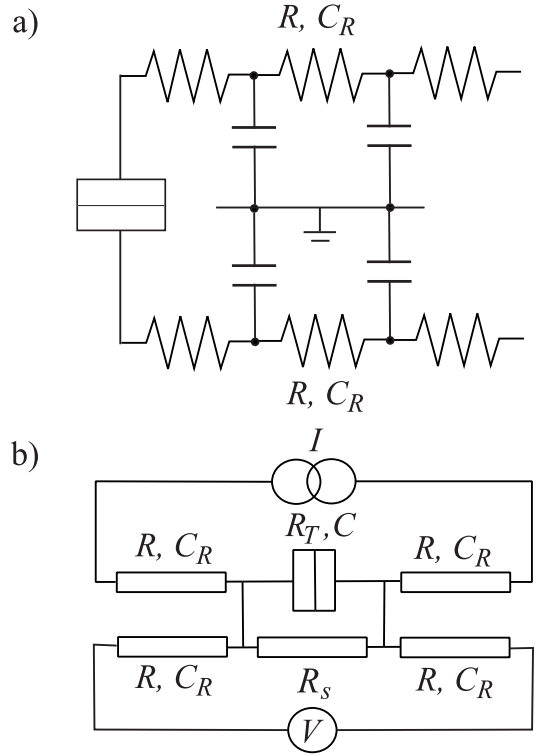}
\end{center}
\caption{a) High frequency model for resistive measurement leads.
Microstrip leads are viewed as transmission lines with the total
capacitance $C_R$ and resistance $R$ per each.  b) Measurement setup.
}
\label{f1}
\end{figure}

At high frequencies resistors in the circuit cease to be lumped elements
and
should be considered as transmission lines with distributed resistance,
capacitance, and inductance.
The admittance of a double transmission line shown in Fig. \ref{f1}
(drawn without inductive elements)  is
\begin{equation}
Y_L (\omega)  ={1 \over 2}\sqrt{ i\omega C_R \over R-i\omega L}
\cot{\sqrt{i\omega  C_R(R-i\omega L)  }},
      \label{trans-ad}\end{equation}
where $R$, $L$, and $C_R$ are the total resistance, inductance, and
capacitance of an individual transmission line. At low frequency
($\omega
\ll 1/RC_R,~R/L$), the admittance becomes purely ohmic,
{\it i.e.}, it behaves as a lumped ohmic resistor: $Y_L \approx 1/2R$.

For a low-resistance transmission line with $R \ll \sqrt{L/C_R}$, the
resistance may be neglected, but in the high-frequency limit
the double transmission line behaves, nevertheless, as an ohmic resistor with
a real impedance $2\sqrt{L/C_R}$ (the energy is lost via
radiation along an infinite transmission line). Then the voltage tail is
$\propto 1/V$
as for a lumped resistor:
\begin{equation}
V_t= {R_K \over 2}\sqrt{C_R\over L} \left(\frac{e}{2\pi C }\right)^2
{1\over
V}  ~.
 \label{IV-lr}  \end{equation}
But for the high-resistance  line with $R \gg \sqrt{L/C_R}$ one may
neglect inductance, and the real part of the admittance in the
high-frequency limit becomes
\begin{equation}
\mbox{Re} Y_L(\omega) \approx {1\over 2} \mbox{Re} \sqrt{-i\omega C_R
\over
R} ~.
    \end{equation}
Then the voltage tail decreases slower, as $1/\sqrt{V}$:
\begin{eqnarray}
V_t &=& {a_{1/2} \over V^{1/2}} \nonumber \\
a_{1/2} &=& {1 \over 3 \pi^{3/2}} \sqrt{R_K
C_R \over RC}\left(e \over C\right)^{3/2} ~.
    \label{iv}
\end{eqnarray}
But irrespective of the magnitude of the dissipative component, at very
high frequency  $\omega
\gg {R/L}$ (high voltage $V \gg (\hbar/e){R/L}$)) the (double) 
transmission line
becomes again ohmic with the real impedance $Z_L \approx 2
\sqrt{L/C_R}$ which is much smaller than $R$ if the line is long (since
$R$ is
proportional to the line length, but $L/C_R$ is not).  However, if $R$
is
large enough, this happens for voltages too high to be relevant in
the experiments. Altogether, both high-voltage tails, Eqs. (\ref{IV-lr})
and (\ref{iv}), become valid when the ``uncertainty'' time
 $\tau_V=\hbar/eV$ introduced by Nazarov \cite{N} becomes less than the
relaxation time $RC_R$ of the circuit.

\subsection{Stray capacitance and the horizon model}

One way to describe the high-frequency effect of environment
is to use
the so-called  ``horizon model'' \cite{WDH,Pekola2}. It represents the
effect of the environment as due to stray capacitance of leads described
by transmission lines.
The relevant stray capacitance originates from the length of
the transmission line over which an electromagnetic signal from the
junction
can travel during the uncertainty time  $\tau_V=\hbar/eV$. This length
is
called the ``horizon'' length $v_{ph} \tau_V$. Here $v_{ph}$ is  the
velocity of the signal propagation.  Indeed, the effective voltage
dependent
capacitance incorporating the effect of the voltage tail
(see Eqs.~(\ref{IVc},\ref{ohmic})) can be
presented as
\begin{equation}
\tilde C =C + {C^2 \over e}V_t(V) = C + c_R v_{ph} \tau_V ~,
   \label{c-R} \end{equation}
where $c_R=C_R/{\cal L}$ is the capacitance per unit length and ${\cal
L}$
is the length of the transmission line. In the past
\cite{WDH,Pekola2} the horizon model was used for lossless
(low-resistance) transmission lines when
\begin{equation}
\tilde C =C + {1\over 2 \pi^2} \sqrt{C_R \over L} {h \over eV}
          \end{equation}
and $v_{ph}=1/\sqrt{c_R l}$ is of the order of the light velocity (here
$l=L/{\cal L}$ is the inductance per unit length). But the model works
also for lossy (high-resistance) lines when
\begin{eqnarray}
\tilde C = C + {4 \over 3 \pi^{3/2}} \sqrt{R_K C_R \over RC}\left(e
C\right)^{1/2}{1\over V^{1/2}} \nonumber \\
= C + {4 \over 3 \pi^{3/2}} \sqrt{h C_R \over ReV}~.
    \label{corC} \end{eqnarray}
Comparing it with Eq. (\ref{c-R}) one sees that $v_{ph} \sim
\sqrt{\omega / c_R r} \sim \sqrt{ eV\ /\hbar c_R r}$, and again on the
order
of the phase velocity along the transmission line (here $r=R/{\cal L}$),
but for the lossy line this velocity is frequency(voltage)-dependent and
much less than the speed of light. Thus, the horizon picture presents a
good
qualitative picture of the effect of environment on the high-voltage
asymptotics.

Originally the horizon model was introduced by B\"uttiker and Landauer
\cite{BL}, who assumed the characteristic time to be the traversal time
of
tunneling which is extremely small (about 10$^{-15}$ sec). In fact,
B\"uttiker and Landauer \cite{BL} considered the process of
tunneling itself which indeed can be affected by the circuit only on
such
short time scales. In the phase-correlation theory  \cite{Ingold}
the circuit noise influences the phase factor in the tunneling
Hamiltonian
({\it i.e.} the
phases of the quantum states on both sides of the junction),
but not the tunneling amplitude itself which is  characterized by
constant
junction conductance $1/R_T$. This influence is possible
over a distance which is the horizon length determined by the
uncertainty
time $\hbar/eV$.

\section{Experiment} \label{Exp}

\subsection{Description of samples}

Our sample consists of a aluminum tunnel junction (area 150*150 nm$^2$),
shunted or unshunted,
connected to four measurement
leads via thin film Cr resistors (25 $\mu $m long) that are
located
within 3 $\mu$m from the junction (see inset of Fig.~2).
The circuits were fabricated
using electron beam lithography and triple-angle evaporation on top
of oxidized silicon substrate (SiO$_2$ thickness $\sim$ 100 nm).
The tunnel barriers were formed by oxidizing the bottom aluminum
electrode in $O_2$ at 0.1 mbar for 5 minutes.
The Cr resistors and shunt (10-15 nm thick, 100 nm wide) were
evaporated
at right angle of incidence. An accurate dosage of resistor wires
ensured that the Al replicas were evaporated on the side of the resist
and thus removed during lift-off.
The shunt resistances $R_s$, made of a 3-6 $\mu $m section of Cr,
varied between 4 and 22 k$\Omega$;
this value was deduced using the length
of the shunt and the measured resistivity $r$ of the Cr
sections in the leads.
In order to compare the highly resistive Cr samples with those in
low impedance
environment, we have fabricated a reference junction consisting of thick
Al
leads. See Table I for description of samples with different
environments.

On the dilution refrigerator, the samples were
mounted inside a tight copper enclosure and the measurement leads were
filtered
using 0.5 m of Thermocoax cable.

\subsection{Effective electric circuit and the fitting
formula}

The elements of the effective electric circuit in the experiment are the
tunnel junction itself which is a lumped element with resistance $R_T$
and
capacitance $C$, four leads with resistance $R$ and stray
capacitance $C_R$ each, and shunt resistance $R_s$ (see Fig.~1b).
An estimation shows that for voltage interval
studied by us both the leads and the shunt are in the high-voltage
regime
where they must be considered as transmission lines: lossy lines for a
shunt
and leads made from Cr which produce the asymptotic square-root law
($1/\sqrt{V}$ -tails), while  lossless lines for Al leads contribute  to
the circuit noise as pure ohmic elements (the $1/V$-tail). Thus, both
$1/V$- and
$1/\sqrt{V}$-tails are present simultaneously in our fitting in high
voltage
regime ($V > k_BT/e, e/C$) which is based on the formula
\begin{equation}
I={V\over R_s} + {1\over R_T}\left(V -{e \over 2C} + {A_1 \over V}
+{A_{1/2}
\over \sqrt{V}} \right) + g V^3~.
      \label{IV}  \end{equation}
By introducing a cubic term $g V^3$ into the fit we take into account
the
nonlinear background at large voltages. Even at large voltages
($\sim$10 mV)
the strength of the cubic background
does not exceed the total contribution from
the power law tails. This makes it possible to resolve the power law
dependence of the tail. In our measurement
scheme, a small ac excitation can be added to bias current sweep
to directly measure the differential conductance $dI/dV$.  The
fitted $g$ agrees with the parabolic background in the differential
conductance measurements.

The values of parameters $A_1$ and $A_{1/2}$  expected  from the
theory are:

(i) Unshunted junction with low impedance Al leads
\begin{eqnarray}
A_1 &=& R_K  \left(e \over 2\pi C \right)^2
\left({1 \over R_T} +\sqrt{C_R \over L} \right),   \nonumber \\
A_{1/2} &=& 0~.
       \label{A} \end{eqnarray}

(ii) Unshunted junction with resistive Cr leads
\begin{eqnarray}
A_1 &=& {R_K \over R_T} \left(e \over 2\pi C \right)^2~,
   \nonumber \\
A_{1/2} &=& 2 a_{1/2}~.
       \label{A2} \end{eqnarray}

(iii) Shunted junction with resistive Cr leads
\begin{eqnarray}
A_1 &=& {R_K \over R_T} \left(e \over 2\pi C \right)^2~,
   \nonumber \\
A_{1/2} &=& 3a_{1/2}~.
       \label{A3} \end{eqnarray}

The factors 2 and 3 in Eqs. (\ref{A2}) and (\ref{A3}) appear because
four
leads are equivalent to two double transmission lines shown in Fig.~1.
Thus the total
number of double transmission lines is 2 and 3 for the unshunted and the
shunted
case, respectively.\footnote{The shunt is considered as a two separate
sections with resistance
$R_s/2$ each.} The shunt resistance $R_s$ is known
whereas $R_T$ and the capacitance $C$ as well as the parameter $A_1$ or
$A_{1/2}$ are fitted to the $IV$ curve. In the case of lossless Al line
(i)
the parameter $A_1$ is fitted freely since it depends not only on $R_T$,
but
also on the admittance
$\sqrt{C_R/L}$. In the case of Cr leads (ii,iii) $A_1$ is fixed by
$R_T$ and
$C$ and the parameter $A_{1/2}$ is fitted.

In the fabrication process the parameters (electron dose) are equal both
for
the shunt and the leads. This means that the capacitance and resistance
per
unit length are equal for Cr shunts and leads. The capacitance per unit
length may be estimated from that of a prolate ellipsoid \cite{LL}:
\begin{equation}
c_R=4\pi \epsilon_0 \epsilon_{eff} \frac{\sqrt{1-(b/a)^2}}
{{\mathrm{ln}}(a/b+\sqrt{(a/b)^2-1})},
\label{Cest}
\end{equation}
where $a$ and $b$ are the larger and smaller radii of the ellipsoid, 
respectively. These are related to the length $l$, thickness $t$, and
width $w$ of the Cr lead through $b \sim \sqrt{tw} \sim$ 30 nm, 
and $a \sim l \sim $ 10 $\mu$m. For our silicon substrate
$\epsilon_{eff} \approx 6$.
Using these values we get $c_R$=100
aF/$\mu$m for our typical Cr environment.
Note that the capacitance per unit length
given by Eq. (\ref{Cest}) depends weakly on the length of the line.
The fit results are compared to
theoretical estimates in Table II. Using the estimated capacitance
$c_R$=100
aF/$\mu$m and the measured resistivity 4 k$\Omega$/$\mu$m, one obtains
for
the attenuation coefficient $\sqrt{\omega rc_R/2} \sim$ 0.57 $\mu$m$^{-1}$ at
1 mV ($\omega = 1.6
\cdot 10^{12}$ 1/s ). Hence,
even though our Cr resistors are rather short, they can be viewed
practically as infinite.

The Al leads must be characterized by the microstrip impedance
\begin{equation}
\sqrt{l/c_R} = \frac{Z_0}{2\pi \sqrt{\epsilon_{eff}}} {\mathrm{ln}}(8h/w)
\equiv Z_0^{eff},
\end{equation}
where $Z_0$ is the free-space impedance of 377 $\Omega$, $h \sim$
600 $\mu$m is the distance
from the ground plane, and $w \sim$ 200 nm is the width of the strip.
This yields impedance of 260 $\Omega$ corresponding
to capacitance per unit length of 30 aF/ $\mu$m.

\subsection{Experimental results}

For  presentation,
it is convenient to subtract off the linear part $I=V/R_{tot}$
with $R_{tot}^{-1}=R_T^{-1}+R_s^{-1}$. Thus we can plot the
``excess'' current
\begin{equation}
I_e  =  V/R_{tot}-I+g V^3  =  \frac{1}{R_T} \left(\frac{e}{2 C}
-\frac{A_1}{V}-\frac{A_{1/2}}{\sqrt{V}}\right)
\end{equation}
as a function of voltage $V>k_BT/e$.

\begin{figure}
\begin{center}
\epsfxsize=75mm \epsffile[66 330 584 714]{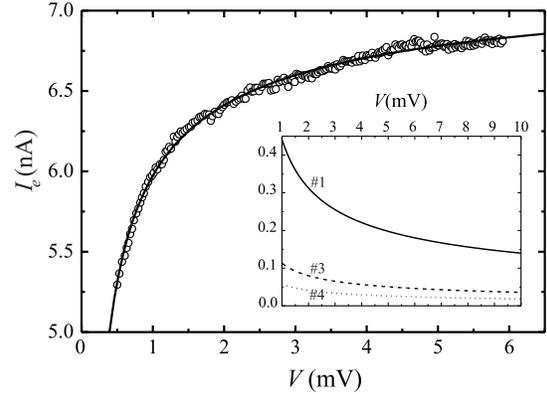}
\end{center}
\caption{Reduced ``excess'' current $I_e$
vs. voltage $V$ for sample 3 with tunneling
resistance $R_T$=11.1 k$\Omega$ and a resistive 0.1x10 $\mu$m Cr
shunt ($R_s$=22.4 k$\Omega$). Solid line illustrates fit using lossy RC
transmission line formulas Eq. (\ref{A3}).
Inset shows the ratio between two tail contributions arising from
tunnel junction and transmission line.}
\label{f2}
\end{figure}

Figure \ref{f2} shows an example of an $I_e$ vs. $V$-curve
measured on a junction with
$r$=4 k$\Omega / \mu$m Cr leads (sample 3). The best fit is obtained
with the
lossy RC-line formula of Eq. (\ref{A3})
with $1/\sqrt{V}$-tail. This
yields $c_R$=210 aF/$\mu$m for the specific capacitance of the
Cr leads, deviating by a factor of two from the estimated value of
100 aF/$\mu$m of Eq. (\ref{Cest}). The results on $1/\sqrt{V}$-tails were
found to be independent of
temperature in the range 0.1 - 1 K. This is in agreement with theory
(see Sec. II A),
according to which the tails depend
only on the quantum part of Johnson-Nyquist noise.

\begin{figure}
\begin{center}
\epsfxsize=75mm \epsffile[50 237 450 544]{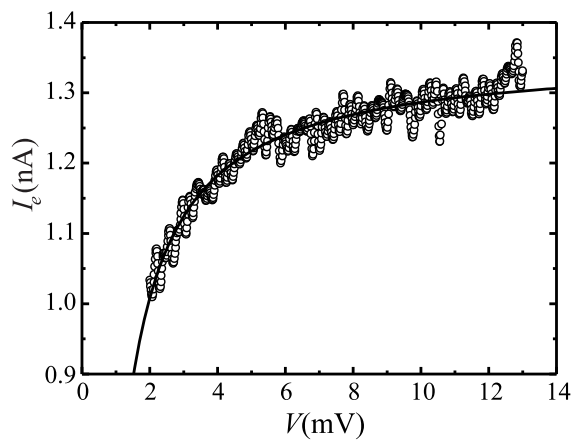}
\end{center}
\caption{``Excess'' current vs. voltage for an unshunted full Al sample
with tunneling resistance $R_T$=76.0 k$\Omega$.
Fit using lossless transmission line formulas Eq. (\ref{A}).
}
\label{f3}
\end{figure}

Figure \ref{f3} shows an $I_e$ vs. $V$ -curve of unshunted sample 2
with thick aluminum leads. As expected, the lossless transmission line
formula (\ref{A}) fits with impedance $\sqrt{L/C_R} \approx $ 570 $\Omega$.
This number agrees with Wahlgren {\it{et al}}. \cite{WDH} who calculated
$R_{env}$= 440 $\Omega$ from the low-voltage data of similar unshunted
single tunnel junction.

The ratio between the tails $A_1/V$ and $A_{1/2}/\sqrt{V}$ is plotted
in the inset of Figure \ref{f2}. Even if the magnitude
$A_{1/2}/\sqrt{V}$ is always larger than $A_1/V$, their
relative magnitude depends on
the tunnel resistance $R_T$. In the strong tunneling regime $R_T <
R_K$,
$A_1/V$ becomes more dominant than in the samples with larger $R_T$.
This means that one must take strong tunneling effect into
account
investigating the high voltage tails, and for this it is sufficient
to include the ohmic tunneling resistance $R_T$ into the effective circuit
for calculation of the Johnson-Nyquist noise.

\begin{figure}
\begin{center}
\epsfxsize=75mm \epsffile[52 430 470 733]{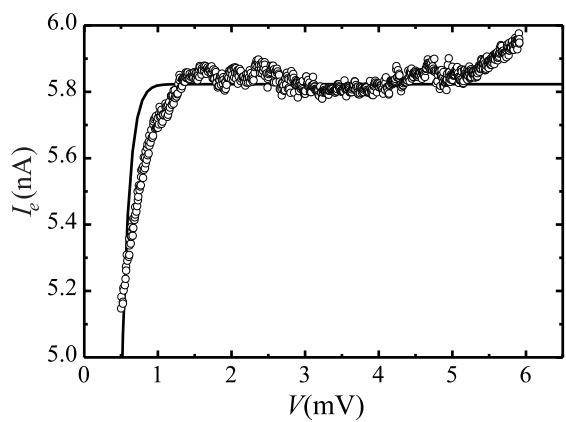}
\end{center}
\caption{Fit to sample 3 with $R_s$=22.4 k$\Omega$ using voltage-
fluctuation theory. The background cubic
nonlinearity used in the fitting formula was orders of magnitude
larger than the exponential tail of Eq. (\ref{IV-l}).
}
\label{f4}
\end{figure}

Figure \ref{f4} shows a fit using voltage-fluctuation theory Eq.
(\ref{IV-l}).
As well as in the other fits, a nonlinear background
$g V^3$ was needed in the fitting. However, even the best fit
yields a magnitude
for the exponential tail which is orders of magnitude smaller than the
cubic background. This is an evidence against the
voltage-fluctuation model.

Table II summarizes the fit results by showing the fitted $R_T$, $C$,
and the parameter $\sqrt{R_K C_R / RC}$ proportional to the fitting
parameter $A_{1/2}$. The theoretical value is obtained from calculated
value of capacitance per unit length Eq.(\ref{Cest}).
The reference sample with lossless Al leads is characterized by
the dimensionless impedance $R_K \sqrt{C_R / L}$ comparable
to $R_K / Z_0^{eff}$.

\section{Concluding discussion}

We have studied experimentally high-voltage asymptotics of the $IV$
curves
for small normal tunnel junctions and
detected power-law voltage tails of $IV$ curve when
approaching the linear law $V=IR +e/2C$. Our data are in a good
agreement with theoretical predictions of the quantum theory of
environment (the phase-correlation theory). Despite some
numerical-factor
discrepancy for high-resistance Cr leads which may be ascribed to
inaccuracy of our effective circuit, the voltage tail grows with
increasing
lead admittance as predicted by the quantum theory.

Voltage tails of the form $1/V$,
typical for low-resistance leads, were experimentally studied and
discussed
by Wahlgren {\it et al.} within the horizon picture \cite{WDH}. We have
detected
slower voltage tails $1/\sqrt{V}$ using high-resistance chromium leads.
We
have shown that the horizon model provides a good qualitative picture
also
for this type of environment, but only if one takes into account that
the
high-resistance leads behave as lossy transmission lines in which the
electromagnetic signal travels with a frequency-dependent velocity
that is much less than the velocity of light.

The effect of environment is in fact a result of Johnson-Nyquist noise
in the electric circuit. It is important that the power-law
tails at high voltages are connected only with the quantum part of
Johnson-Nyquist
noise. Note that in order to avoid thermal noise when
studying the low-voltage part
of the $IV$ curve, the condition $k_B T \ll e^2/C$ must be satisfied.
For the
high-voltage tails the condition may be much weaker {\it viz.} $ k_B T
\ll eV/\pi$
which is well satisfied in our experimental studies. Thus,
detection of these tails in a good agreement with theory is a rather
unique verification
of quantum zero fluctuations in macroscopic systems.

Our experimental conditions included the case of strong tunneling when
the
junction resistance $R_T$ was less than the quantum resistance $R_K$.
The strong-tunneling corrections to the
environmental modes at high voltages
can be simply incorporated by
including the junction resistance $R_T$ into the effective electric
circuit for calculation of the quantum noise.

\section*{Acknowledgements}

We acknowledge interesting discussions with G.-L. Ingold and A.D.
Zaikin.
This work was supported by the Academy of Finland
and by the Human Capital and Mobility Program ULTI of the European
Union.

\begin{table}
\caption{Measured shunted junctions.
The value of $R_s$ is estimated from the known wire resistivity $r$.}

  \begin{tabular}{c|c|c}
    sample & \hspace{-12mm} $R_s(k\Omega )$ & \hspace{-12mm} Description
of environment \\
   \hline 1 & \hspace{-12mm} $\infty$ & Cr leads only, $r$=4 k$\Omega/
\mu m$ \\
   \hline 2 & \hspace{-12mm} $\infty$ & Al leads only  \\
   \hline 3 & \hspace{-12mm} 22.4     & Cr shunt/leads, $r$=4
k$\Omega/ \mu m$ \\
   \hline 4 & \hspace{-12mm} 4.2      & Cr shunt/leads, $r$=1.5
k$\Omega/ \mu m$ \\
  \end{tabular}
\end{table}

\begin{table}
\caption{Fit results. $R_T$ and $C$ are fitted to the data.
Parameters $\sqrt{R_K C_R / RC}$ and $R_K \sqrt{C_R/L}$ are
calculated from the fitted parameters $A_{1/2}$ and $A_1$,
respectively.
These are compared with theoretical estimates (see text).}

\begin{tabular}{c|l|l|l|l}
     sample & $R_T$(k$\Omega)$ & $C$(fF) &
      \multicolumn{2}{l}{$\sqrt{R_K C_R / RC}$} \\
     &      &     & fit   & theory \\
   \hline
   1 & 4.3  & 1.1 & 1.11  & 0.77 \\
   \hline
   3 & 11.1 & 1.0 & 1.16  & 0.80 \\
   \hline
   4 & 18.4 & 0.1 & 3.84  & 4.15 \\
   \hline \hline
     sample & $R_T$(k$\Omega)$ & $C$(fF) &
      \multicolumn{2}{l}{$R_K \sqrt{C_R / L}$} \\
     &      &     & fit   & $R_K/Z_0^{eff}$ \\
   \hline
   2 & 76.0 & 0.7 & 45.3   & 99 \\
 \end{tabular}
\end{table}

\end{document}